\global\def\draftcontrol{0}

   \def\versionno{ ksbh}

\catcode`\@=11

\expandafter\ifx\csname draftcontrol\endcsname\relax\global\def\draftcontrol{0}
\fi

{\count255=\time\divide\count255 by 60
\xdef\hourmin{\number\count255}
\multiply\count255 by-60\advance\count255 by\time
\xdef\hourmin{\hourmin:\ifnum\count255<10 0\fi\the\count255}}
\def\draftdate{\number\month/\number\day/\number\year\ \ \ \hourmin }

\newcommand\makepapertitle{\par
  \begingroup
    \renewcommand\thefootnote{\@fnsymbol\c@footnote}%
    \def\@makefnmark{\rlap{\@textsuperscript{\normalfont\@thefnmark}}}%
    \long\def\@makefntext##1{\parindent 1em\noindent
            \hb@xt@1.8em{%
                \hss\@textsuperscript{\normalfont\@thefnmark}}##1}%
     \newpage
     \global\@topnum\z@   
     \@makepapertitle
     \thispagestyle{empty}\@thanks
  \endgroup
  \setcounter{footnote}{0}%
  \global\let\thanks\relax
  \global\let\makepapertitle\relax
  \global\let\@makepapertitle\relax
  \global\let\@thanks\@empty
  \global\let\@author\@empty
  \global\let\@date\@empty
  \global\let\@title\@empty
  \global\let\title\relax
  \global\let\author\relax
  \global\let\date\relax
  \global\let\and\relax
  \def\version{\let\version\@version\@gobble}
}
\def\@makepapertitle{%
  \newpage
   \ifnum\draftcontrol=1 {}
   \version\versionno
   \vskip 3em%
   \else
   \hfill\hbox to 3cm {\parbox{4cm}{\@pubnum}\hss}%
   \vskip 3em%
   \fi
   \begin{center}%
   \let \footnote \thanks
     {\LARGE {\@title}}%
     \vskip 1.5em%
     {\normalsize
       \lineskip .5em%
       \begin{tabular}[t]{c}%
         \@author
       \end{tabular}\par}%
     \vskip 1.5em%
     {\@bstract}%
     \end{center}%
     \vskip 1.5em
     \@date%
   \par
}

\gdef\@pubnum{}
\def\pubnum#1{%
  \gdef\@pubnum{#1}}

\gdef\@bstract{}
\def\Abstract#1{%
  \gdef\@bstract{%
   \parbox{\textwidth-0pc}{%
   \centerline{\bf Abstract}\penalty1000%
\kern.2cm%
\noindent
\renewcommand\baselinestretch{1.0}%
{#1}}}
}

\def\ps@paper{\let\@mkboth\@gobbletwo%
     \ifnum\draftcontrol=1
    \def\@oddfoot{\hbox to \textwidth{\tiny \versionno \hfil\tiny\draftdate}%
    \hskip -\textwidth \hbox to \textwidth{\hfil\rm\thepage\hfil}}%
     \else\def\@oddfoot{\hbox to \textwidth{\hfil\rm\thepage\hfil}}
     \fi
     \let\@evenfoot\@oddfoot
}

\def\body{\clearpage
          \pagestyle{paper}
    }

\def\@version#1{\ifnum\draftcontrol=1
\typeout{}\typeout{#1}\typeout{}
\vskip3mm\centerline{\hbox{\fbox{\normalsize{\tt DRAFT -- #1 -- }
                   {\draftdate}}}}\vskip3mm
\fi}
\let\version\@version
\long\def\eqlabel#1{\ifnum\draftcontrol=1
                    \tag@false  
                    \tag*{(\theequation) \hbox to -0.2cm{\hspace{0cm}\small{#1}\hss}}
                    \refstepcounter{equation}
                    \edef\@currentlabel{\theequation}
                    \ltx@label{#1}          
                    \else
                    \label{#1}
                    \fi
                    }
\let\st@bibitem\@bibitem
\let\st@lbibitem\@lbibitem
\ifnum\draftcontrol=1
  \def\@bibitem#1{%
    \st@bibitem{#1}\a@@label{#1}\ignorespaces}
  \def\@lbibitem[#1]#2{%
    \st@lbibitem[#1]{#2}\a@@label{#2}\ignorespaces}
  \def\a@@label#1{%
    \gdef\a@lab{\smash{\normalfont\small#1}}
    \ifvmode
      \if@inlabel
        \global\setbox\@labels\hbox{%
          \llap{\a@lab\let\a@lab\relax
                \kern\@totalleftmargin\kern\marginparsep}%
          \box\@labels}%
      \fi
    \fi}
\fi

\documentclass[12pt,letterpaper]{article}

\usepackage{amsmath,amssymb,array,calc,epsfig,rotating,bm}
\usepackage[sort]{cite}
\usepackage{graphicx}
\usepackage{psfrag,verbatim}
\usepackage{xcolor}

\ifnum\draftcontrol=1
\fi

\renewcommand\baselinestretch{1.25}
\renewcommand\section{\@startsection {section}{1}{\z@}%
                                   {-3.5ex \@plus -1ex \@minus -.2ex}%
                                   {2.3ex \@plus.2ex}%
                                   {\normalfont\large\bfseries}}
\renewcommand\subsection{\@startsection{subsection}{2}{\z@}%
                                   {-3.25ex\@plus -1ex \@minus -.2ex}%
                                   {1.5ex \@plus .2ex}%
                                   {\normalfont\normalsize\bfseries}}
\renewcommand\subsubsection{\@startsection{subsubsection}{3}{\z@}%
                                   {-3.25ex\@plus -1ex \@minus -.2ex}%
                                   {1.5ex \@plus .2ex}%
                                   {\normalfont\normalsize\it}}
\renewcommand\paragraph{\@startsection{paragraph}{4}{\z@}%
                                   {-3.25ex\@plus -1ex \@minus -.2ex}%
                                   {1.5ex \@plus .2ex}%
                                   {\normalfont\normalsize\bf}}

\numberwithin{equation}{section}

\def\revise#1       {\raisebox{-0em}{\rule{3pt}{1em}}%
                     \marginpar{\raisebox{.5em}{\vrule width3pt\
                     \vrule width0pt height 0pt depth0.5em
                     \hbox to 0cm{\hspace{0cm}{%
                     \parbox[t]{4em}{\raggedright\footnotesize{#1}}}\hss}}}}

\newcommand\nxt[1]  {\\\fnxt#1}
\newcommand{\ie}{{\it i.e.,}\ }

\def\cale         {{\cal E}}
\def\calf         {{\cal F}}

\def\calm         {{\cal M}}
\def\caln         {{\cal N}}
\def\calo         {{\cal O}}
\def\calp         {{\cal P}}

\def\complex      {{\mathbb C}}

\def\zet          {{\mathbb Z}}

\def\tr           {\mathop{\rm Tr}}

\def\sqr#1#2{{\vcenter{\vbox{\hrule height.#2pt
 \hbox{\vrule width.#2pt height#1pt \kern#1pt
 \vrule width.#2pt}\hrule height.#2pt}}}}

\def\csb{\chi SB}

\def\a{\alpha}

\def\e{\epsilon}

\def\g{\gamma}

\def\aa1{\phi}
\def\cc1{\psi}

\def\Om{\Omega}
\def\om{\Omega}

\def\csb{{\chi\rm{SB}}}

\catcode`\@=12

\begin{document}


\title{\bf Klebanov-Strassler black hole}

\date{December 8, 2018}

\author{
Alex Buchel\\[0.4cm]
\it Department of Applied Mathematics,\\
\it Department of Physics and Astronomy\\ 
\it University of Western Ontario\\
\it London, Ontario N6A 5B7, Canada;\\
\it Perimeter Institute for Theoretical Physics\\
\it Waterloo, Ontario N2J 2W9, Canada\\[0.4cm]
}

\Abstract{We construct a black hole solution on warped deformed
conifold in type IIB supergravity with fluxes. The black hole has
translationary invariant horizon and is a holographic dual to a
thermal homogeneous and isotropic state of a cascading $SU(K+P)\times
SU(K)$ $\caln=1$ supersymmetric gauge theory with spontaneously broken
chiral symmetry.  We discuss thermal properties of the new black hole
solutions.  We comment on implications of the new black hole solutions
for the landscape of KKLT de Sitter vacua in string theory.
}

\makepapertitle

\body

\version\versionno
\tableofcontents

\section{Introduction}
A conifold is a complex 3-dimensional manifold described by the following equation
in $\complex^4$, 
\begin{equation}
\sum_{n=1}^4\ z_n^2=0\,.
\eqlabel{conmetric}
\end{equation}
Owing to explicitly known  Ricci-flat metric \cite{Candelas:1989js}, the conifold appeared prominently
in string theory and supergravity:
\nxt In a holographic context \cite{m1}, placing a large number $K$ of $D3$ branes
at the tip of the (singular) conifold realizes a duality for $\caln=1$ $SU(K)\times SU(K)$
superconformal gauge theory, also known as Klebanov-Witten gauge theory \cite{Klebanov:1998hh}.
Further wrapping $P$ $D5$ branes on a 2-cycle of a conifold realizes a correspondence
with $\caln=1$ $SU(K+P)\times SU(K)$ Klebanov-Strassler (KS) gauge theory \cite{Klebanov:2000hb}.
KS gauge theory undergoes an infinite sequence - a {\it cascade} - of self-similar Seiberg duality
\cite{Seiberg:1994pq} transformations in the UV; it confines with the spontaneous chiral symmetry
breaking in the IR. 
\nxt In \cite{Giddings:2001yu} (GKP) it was pointed out that type IIB string
theory compactified on warped throat geometries with fluxes (local version of which
is precisely that of KS gauge theory gravitational dual) produces no-scale
$\caln=1$ supersymmetric Minkowski vacua,  naturally generating large hierarchies of physical scales.
GKP compactifications fix all complex structure moduli, leaving at least a single K\"ahler modulus
(an overall volume of a compact 6-dimensional manifold) unfixed. KKLT \cite{Kachru:2003aw}
further argued that non-perturbative corrections in GKP compactifications
fix the overall volume K\"ahler modulus leading to $\caln=1$ SUSY preserving $AdS_4$ vacua.
Adding $\overline{D3}$ to compactifications lifts AdS vacua to de Sitter. KKLT construction
has been taken as the primary evidence for a landscape of de Sitter vacua in string theory.

In this paper we present new results regarding black hole solutions in type IIB supergravity
on warped deformed conifold with fluxes. Our results are relevant both
for holography and the landscape:
\begin{itemize}
\item Black hole geometries in holography represent gravitational dual to thermal states of the boundary
gauge theories \cite{Witten:1998zw}. Although the chiral symmetry is spontaneously broken in
the vacuum of KS gauge theory, it is expected to be restored at sufficiently high temperature. This predicts
the existence of black hole solutions dual to thermal states of the cascading gauge theory
plasma with unbroken chiral symmetry \cite{Buchel:2000ch}.
Such black holes resolve the singularity of the Klebanov-Tseytlin \cite{Klebanov:2000nc}
geometry \cite{Buchel:2000ch,Buchel:2001gw,Gubser:2001ri}. KS gauge theory confines in the infrared ---
the holographic dual of this (first-order) phase transition was established
in \cite{Aharony:2007vg}\footnote{To understand the thermodynamics of chirally symmetric states of the
cascading gauge theory plasma it was important to understand the holographic renormalization of the
theory \cite{Aharony:2005zr}.} (ABK). ABK black hole solutions represent the gravitational dual
to deconfined thermal states of KS gauge theory with unbroken chiral symmetry. These black holes cease to exist
below  some critical temperature $T<T_u$
\cite{Buchel:2009bh}, where they join a perturbatively unstable branch (with a negative specific heat  
and condensation of the hydrodynamic sound modes).
ABK black holes are also perturbatively unstable for $T<T_{\csb}=1.00869(0) T_u$
towards development of the chiral symmetry breaking ($\chi$SB) condensates in KS gauge theory plasma   
\cite{Buchel:2010wp}. The end point of the $\csb$ breaking instability in ABK black holes would produce
Klebanov-Strassler black hole --- gravitational backgrounds dual to homogeneous and isotropic
thermal deconfined states of KS gauge theory
plasma with spontaneously broken chiral symmetry. 
Thermodynamics of KS black holes will be discussed in section \ref{hol}.
\item The most controversial aspect of the KKLT construction of de Sitter vacua is the uplift of
$AdS_4$ non-perturbative GKP vacua due to $\overline{D3}$ branes.
The backreacted solutions corresponding to smeared $\overline{D3}$ branes at the tip of the KS solution
were argued to be singular \cite{Bena:2009xk,Bena:2011hz,Bena:2011wh,Bena:2012bk}. Because
the singularity is localized, it must be possible to study it in the local geometry of the GKP background ---
the noncompact gravitational dual to KS gauge theory. The following strategy was proposed in \cite{Bena:2012ek}:
if the singularity due to  $\overline{D3}$ branes at the conifold is physical, it 
should be possible to shield it with a horizon  \cite{Gubser:2000nd}:
\ie there must exist a black hole solution on the conifold
that carries a {\it negative} $D3$ brane charge at the horizon. Black holes with negative $D3$ brane
horizon charge have not been found for a conifold with an unbroken $U(1)$ symmetry
(a gravitational dual to a chiral symmetry in KS gauge theory); neither were found black holes with
negative charge where this $U(1)$ symmetry is broken explicitly \cite{Bena:2012ek}.
KKLT construction requires warped deformed conifolds with spontaneous symmetry breaking $U(1)\supset \zet_2$ ---
thus, one needs to search for negative $D3$ brane charge KS black holes.
We report on this in section \ref{lan}.
\end{itemize}

Constructions of Klebanov-Strassler black holes have been attempted in the past:
the complete  ansatz for the metric and the background fluxes was proposed in
\cite{Buchel:2010wp}\footnote{In later work, \cite{Mia:2012yq}, the authors did not identify all the required thermal
condensates in cascading gauge theory plasma.}. The latter reference  contains the consistent set of equations of motion,
which incorporates as  special cases all previously known solutions
\cite{Buchel:2001gw,Gubser:2001ri,Aharony:2007vg}, dual to thermal states of cascading gauge theory
with unbroken chiral symmetry (the conifold deformation parameter is switched off).
As we already mentioned, the background ansatz of \cite{Buchel:2010wp}
identifies (linearized) perturbative instability of Klebanov-Tseytlin (ABK) black holes at $T< T_{\csb}$
due to fluctuations associated with the conifold deformation parameter --- in the
cascading gauge theory language this is a coupled set of two dimension-3 operators $\calo_3^{1,2}$ and a
dimension-7 operator $\calo_7$, see section 3 of \cite{Buchel:2010wp}. 
Klebanov-Strassler black hole is a solution within the metric ansatz \cite{Buchel:2010wp} with
non-linear thermal expectation values $\calo_3^{1,2}$ and $\calo_7$. It is important to distinguish
black hole solutions where the non-normalizable components of gravitational background fields dual to
$\calo_3^{1,2}$ operators vanish or are nonzero: the genuine Klebanov-Strassler black hole,
dual to thermal deconfined phase of cascading gauge theory with {\it spontaneously} broken
chiral symmetry is the former; black hole solutions with nonzero non-normalizable components
of the fields dual to $\calo_3^{1,2}$ operators describe {\it explicit} breaking of chiral
symmetry by gaugino mass terms. Thermal states of cascading gauge theory
with explicit breaking of chiral symmetry due to gaugino mass terms were extensively studied in
section 4 of  \cite{Buchel:2010wp}\footnote{These computations were independently reproduced
in \cite{Bena:2012ek}.}. It was demonstrated in \cite{Buchel:2010wp} that thermal states in
mass-deformed cascading theory at $T<T_{\csb}$ reduce to chirally-symmetric Klebanov-Tseytlin black holes
in the limit of vanishing gaugino masses, see Fig.~4 in \cite{Buchel:2010wp} --- this was taken
as an evidence that Klebanov-Strassler black holes do not exist. Both a conceptual and a technical
advance allowed a successful construction of Klebanov-Strassler black hole reported in this paper, 18 years since
the first work on a subject \cite{Buchel:2000ch}:
\nxt At a conceptual level, it was realized that KS black holes could never dominated
the canonical ensemble, and thus might not simply exist for $T<T_{\csb}$, in agreement with \cite{Buchel:2010wp}.
For this to be true, the thermal phase diagram of the system should resemble the one
of the "Exotic Hairy Black Holes'' first  discovered in \cite{Buchel:2009ge}. As we will see in section 2.1 below,
this is indeed the case.
\nxt The technical difficulty in numerically constructing KS black holes is the nonzero value of
the thermal expectation value of $\calo_7$ operator. This implies that in numerical solutions
one must keep computational control over the boundary asymptotics (the radial coordinate $r\to\infty$)
of the fields to level $\calo(r^{-7})$, while the leading asymptotic of the fields is $\sim \ln r$.
Furthermore, this thermal expectation value vanishes as $\calo_7\sim |T-T_{\csb}|^{1/2}\to 0$
at the temperature of the spontaneous chiral symmetry breaking $T_{\csb}$. (It is precisely to circumvent
this difficulty,  it was proposed in \cite{Buchel:2010wp} to break chiral symmetry explicitly, and then
construct KS black holes in the limit of vanishing gaugino masses.) In this paper we solved
the technical problem, without introducing gaugino mass terms,
by developing new computational Mathematica scripts to solve relevant differential equations with
arbitrary numerical precision.

In the next two sections we present results of relevance to cascading gauge theory holography
and to the landscape of KKLT de Sitter vacua, omitting all the technical details.
All the necessary technical details are reviewed (following \cite{Buchel:2010wp}) in Appendix.

\section{Holography: phases of the cascading gauge theory}\label{hol}
In our review of the cascading gauge theory and the thermodynamics of its chirally symmetric states we closely follow 
\cite{Buchel:2010wp}.

Klebanov-Strassler gauge theory is  $\caln=1$ four-dimensional supersymmetric $SU(K+P)\times SU(K)$
gauge theory with two chiral superfields $A_1, A_2$ in the $(K+P,\overline{K})$
representation, and two fields $B_1, B_2$ in the $(\overline{K+P},K)$.
This gauge theory has two gauge couplings $g_1, g_2$ associated with 
two gauge group factors,  and a quartic 
superpotential
\begin{equation}
W\sim \tr \left(A_i B_j A_kB_\ell\right)\e^{ik}\e^{j\ell}\,.
\end{equation}
When $P=0$ above theory flows in the infrared to a 
superconformal fixed point, commonly referred to as Klebanov-Witten 
theory. At the IR fixed point KW gauge theory is 
strongly coupled --- the superconformal symmetry together with 
$SU(2)\times SU(2)\times U(1)$ global symmetry of the theory implies 
that anomalous dimensions of chiral superfields $\gamma(A_i)=\gamma(B_i)=-\frac 14$, \ie non-perturbatively large.

When $P\ne 0$, conformal invariance of the above $SU(K+P)\times SU(K)$
gauge theory is broken. It is useful to consider an effective description 
of this theory at energy scale $\mu$ with perturbative couplings
$g_i(\mu)\ll 1$. It is straightforward to evaluate NSVZ beta-functions for 
the gauge couplings. One finds that while the sum of the gauge couplings 
does not run
\begin{equation}
\frac{d}{d\ln\mu}\left(\frac{\pi}{g_s}\equiv \frac{4\pi}{g_1^2(\mu)}+\frac{4\pi}{g_2^2(\mu)}\right)=0\,,
\eqlabel{sum}
\end{equation}
the difference between the two couplings is  
\begin{equation}
\frac{4\pi}{g_2^2(\mu)}-\frac{4\pi}{g_1^2(\mu)}\sim P \ \left[3+2(1-\g_{ij})\right]\ \ln\frac{\mu}{\Lambda}\,,
\eqlabel{diff}
\end{equation}
where $\Lambda$  is the strong coupling scale of the theory and $\g_{ij}$ is an anomalous dimension of operators $\tr A_i B_j$.
Given \eqref{diff} and \eqref{sum} it is clear that the effective weakly coupled description of $SU(K+P)\times SU(K)$ gauge theory 
can be valid only in a finite-width energy band centered about $\mu$ scale:
extending effective description both to the UV 
and to the  IR one necessarily encounters strong coupling in one or the other gauge group factor.
To extend the theory past the strongly coupled region(s) one must perform a Seiberg duality. 
In KS gauge theory a Seiberg duality transformation is a self-similarity transformation of the effective description 
so that $K\to K-P$ as one flows to the IR, or $K\to K+P$ as the energy increases. Thus, extension of the effective 
$SU(K+P)\times SU(K)$ description to all energy scales involves a cascade of Seiberg dualities  where
the rank of the gauge group changes with energy according to  
\begin{equation}
K=K(\mu)\sim 2 P^2 \ln \frac \mu\Lambda\,, 
\eqlabel{effk}
\end{equation}
at least as $\mu\gg \Lambda$.
Although there are infinitely many duality cascade steps in the UV, there is only a finite number of duality transformations as one 
flows to the IR (from a given scale $\mu$). The space of vacua of a generic cascading gauge theory was studied in details in 
\cite{dks}. When $K(\mu)$ is an integer multiple of $P$, the cascading gauge theory confines in the 
infrared with a spontaneous breaking of the chiral symmetry.  

The thermal phase digram of homogeneous and isotropic states in
cascading gauge theory plasma represents competition between three phases:
\begin{itemize}
\item (A): confined phase with spontaneously broken chiral symmetry;
\item (B): deconfined chirally symmetric phase;
\item (C): deconfined phase with spontaneously broken chiral symmetry.
\end{itemize}
Correspondingly, in a dual gravitational description we have:
\begin{itemize}
\item (Ah): thermal KS geometry, \ie Klebanov-Strassler vacuum solution with periodically identified Euclidean
time direction,
\begin{equation}
t_E\ \sim\ t_E+\frac 1T \,.
\eqlabel{tks}
\end{equation}
In this phase all the thermodynamic potentials vanish: the free energy density $\calf$,
the energy density $\cale$ and  the entropy density $s$.
There are nonvanishing condensates of two dimension-3 operators (dual to chiral symmetry
breaking gaugino condensates of both gauge group factors),
and a condensate of a dimension-6 operator \cite{Buchel:2010wp}. 
\item (Bh): Klebanov-Tseytlin black hole. In this phase we have nonvanishing $\{\calf,\cale,s\}$.
There are nonvanishing condensates of two dimension-4 operators, a dimension-6 operator and a dimension-8
operator \cite{Aharony:2007vg}. Condensates of the chiral symmetry breaking operators vanish.
\item (Ch): Klebanov-Strassler black hole. In this phase  we have nonvanishing $\{\calf,\cale,s\}$.
In additional to the condensates present in (Bh), we have condensates of a pair of chiral symmetry breaking dimension-3 operators
(as in (Ah)) and an additional condensate of a dimension-7 operator (also breaking the chiral symmetry) \cite{Buchel:2010wp}.
\end{itemize}
Phase transition between  $A\leftrightarrow B$ is of the first-order \cite{Aharony:2007vg}.

We now turn to a detailed discussion of the phase diagram of the cascading gauge theory plasma.
At temperatures $T\gg \Lambda$ the cascading 
plasma is in the deconfined phase with an unbroken chiral symmetry (B)
\cite{Buchel:2000ch,Buchel:2001gw,Gubser:2001ri}. Here, the temperature-dependent 
effective rank $K(T)$ of the cascading theory is large, compare to $P$ \cite{Aharony:2005zr}:
\begin{equation}
\frac{K(T)}{P^2}=\frac 12 \ln \left(\frac{64\pi^4 }{81}\ \times\ \frac{s T }{ \Lambda^4}\right)\qquad \Longrightarrow\qquad 
 \frac{K(T)}{P^2}\ \approx\ 2\ \ln \frac{T}{\Lambda}\,,\qquad T\gg \Lambda\,.
\eqlabel{kp2}
\end{equation}      
To leading order at higher temperature, the pressure $\calp=-\calf$ and the energy density
$\cale$ are given by \cite{Aharony:2005zr}
\begin{equation}
\begin{split}
\frac{\calp}{sT}=&\frac 14\left(1-\frac{P^2}{K(T)^2}+\calo\left(\frac{P^4}{K(T)^2}\right)\right)\,,\cr
\frac{\cale}{sT}=&\frac 34\left(1+\frac 13\ \frac{P^2}{K(T)^2}+\calo\left(\frac{P^4}{K(T)^2}\right)\right)\,.
\end{split}
\eqlabel{pe}
\end{equation}
At low temperature/energy density we need to distinguish canonical (see fig.~\ref{figure1}) and microcanonical
(see fig.~\ref{figure2})
ensembles.

\subsection{Canonical ensemble}

\begin{figure}[t]
\begin{center}
\psfrag{t}{{$\frac{T}{\Lambda}$}}
\psfrag{f}{{$\hat{\calf}/\Lambda^4$}}
\psfrag{c}{{$\calo_3/\Lambda^3$}}
\includegraphics[width=2.8in]{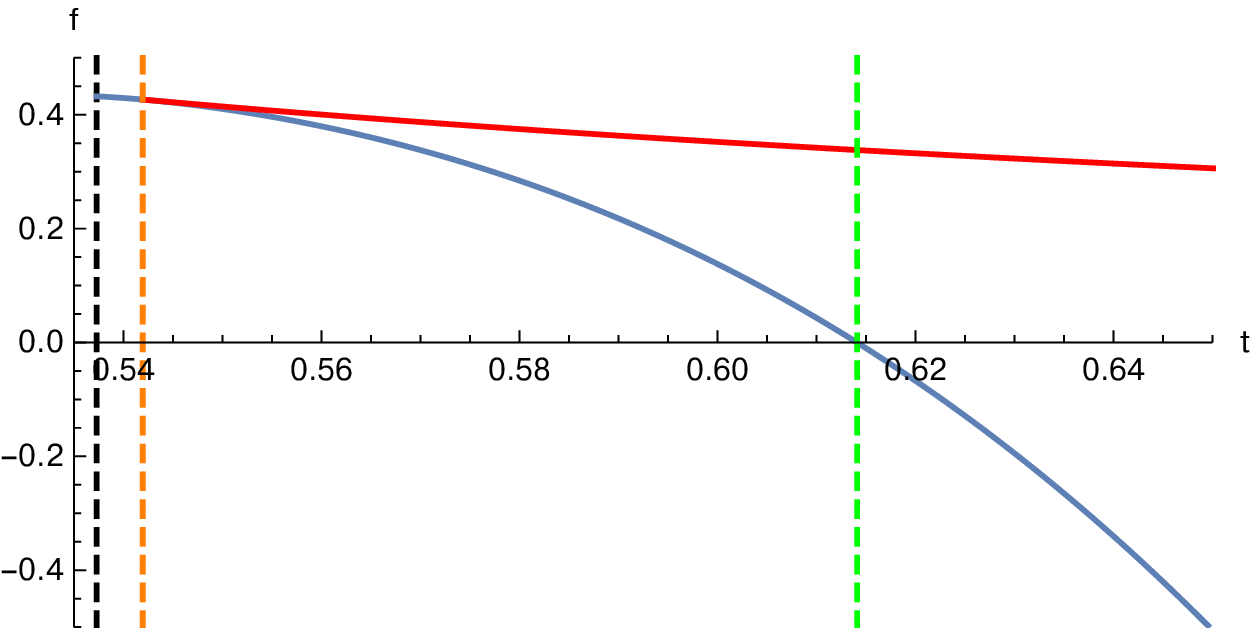}\,\,
\includegraphics[width=2.8in]{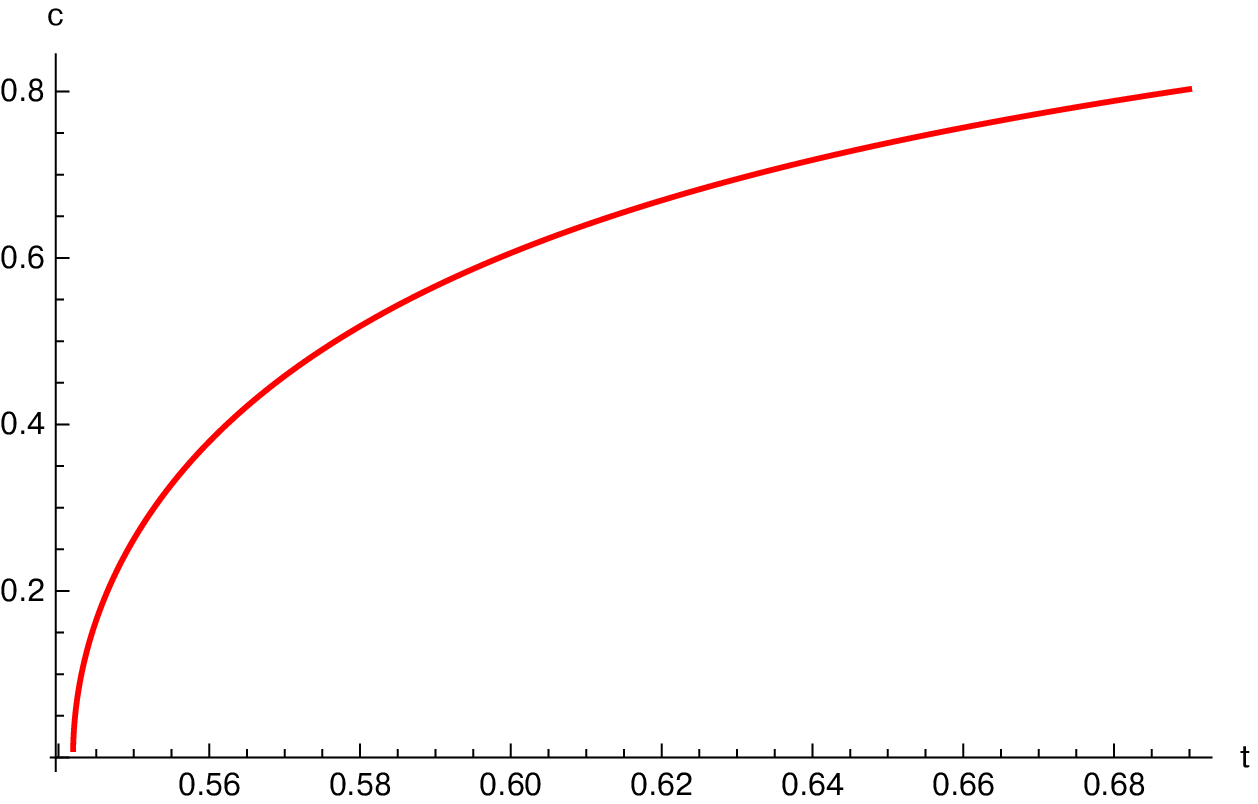}
\end{center}
  \caption{Phase diagram of the cascading gauge theory plasma in canonical ensemble. Solid blue line represents
  deconfined, chirally symmetric phase of the theory --- the gravitational dual to these thermal states is a
  Klebanov-Tseytlin black hole. Solid red lines represent deconfined phase of the theory with the
  spontaneously broken chiral symmetry --- the gravitational dual to these thermal states is a Klebanov-Strassler black hole.
Left panel shows the free energy density versus the temperature of the symmetric and the symmetry broken phases. The vertical dashed green line indicates
the first-order confinement-deconfinement phase transition; the vertical dashed orange line indicates the onset of the perturbative instability
of the chirally symmetric phase; the vertical dashed black line indicates the onset of the instability of
sound waves in chirally symmetric phase of the cascading plasma.
Right panel shows a critical behavior of one of the dimension-3 condensates in the chiral symmetry broken phase of the cascading plasma.  
 } \label{figure1}
\end{figure}

The deconfined chirally symmetry phase (B) (represented by a solid blue curve) extends to temperature \cite{Aharony:2007vg}
\begin{equation}
 T_c=0.6141111(3) \Lambda\,,
\eqlabel{cd}
\end{equation}
below which its free energy density
\begin{equation}
\hat{\calf}=16\pi G_5\ \calf
\eqlabel{deffh}
\end{equation}
becomes positive. This signifies the first-order phase transition to the confined phase with the spontaneously broken
chiral symmetry (A). This phase transition is denoted by a vertical dashed green line. Since  $A\leftrightarrow B$
phase transition proceeds via bubble nucleation, it is non-perturbative. At temperature
\begin{equation}
 T_{\csb}=0.54195(5) \Lambda\,,
\eqlabel{scbt}
\end{equation}
the meta-stable phase (B) becomes perturbatively unstable due to chiral symmetry breaking fluctuations \cite{Buchel:2010wp}.
This instability is denoted by a vertical dashed orange line.
At temperature
\begin{equation}
 T_{u}=0.537286 \Lambda\,,
\eqlabel{u}
\end{equation}
the phase (B) terminates joining a perturbatively unstable branch of
the theory with a negative specific heat \cite{Buchel:2009bh}. The branch with a negative specific heat
has dynamical instability leading to a breakdown of spatial homogeneity in plasma: the sound waves are unstable
\cite{Buchel:2005nt}. The terminal temperature $T_u$ is denoted by a vertical dashed black line.
Note the hierarchy of critical temperatures of homogeneous and isotropic thermal states in cascading plasma:
\begin{equation}
T_u\ <\ T_{\csb}\ <\ T_c\,. 
\eqlabel{thier}
\end{equation}

A natural expectation is that the deconfined phase with spontaneously broken chiral symmetry --- the phase (C) ---
should bifurcate from (B) at $T=T_{\csb}$ where the fluctuations associated with this symmetry breaking
become unstable. The end point of the instability for $T<T_{\csb}$ would be Klebanov-Strassler black hole.
Until this work, the searches for the KSBH were unsuccessful. The right panel of fig.~\ref{figure1} provides
a reason\footnote{Over the years, the author was searching for the KSBH
at $T< T_{\csb}$.}: although the $\csb$ fluctuations are unstable for $T\le T_{\csb}$, the KSBH exists only
for $T\ge T_{\csb}$. In other words, the KSBH is in a class of {\it exotic} black holes originally identified in
\cite{Buchel:2009ge}\footnote{See also \cite{Bosch:2017ccw}.}. The phase (C) of the cascading gauge theory plasma
is denoted by a solid red curve: it has a higher free energy density than the phase (B) at the corresponding
temperature and thus never dominates in macrocanonical ensemble.

\subsection{Microcanonical ensemble}

\begin{figure}[t]
\begin{center}
\psfrag{e}{{$\frac{\hat{\cale}}{\Lambda^4}$}}
\psfrag{s}{{$\hat{s}/\Lambda^3$}}
\psfrag{c}{{$c_s^2$}}
\includegraphics[width=2.8in]{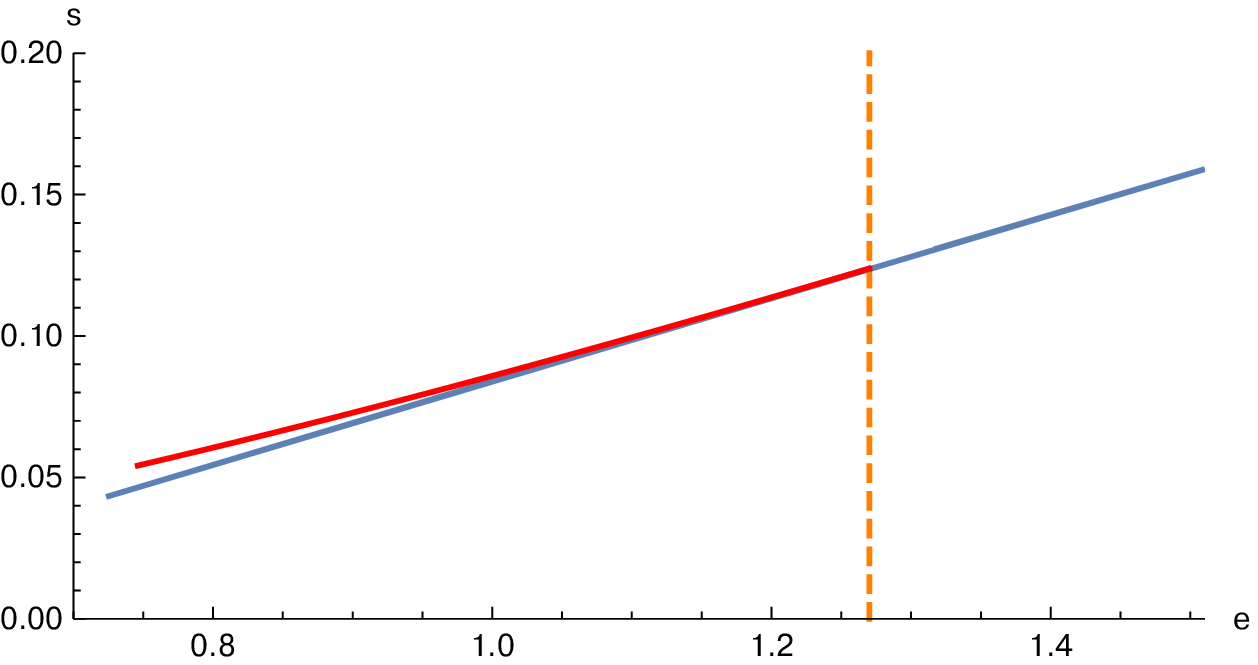}\,\,\,\,\,\,
\includegraphics[width=2.8in]{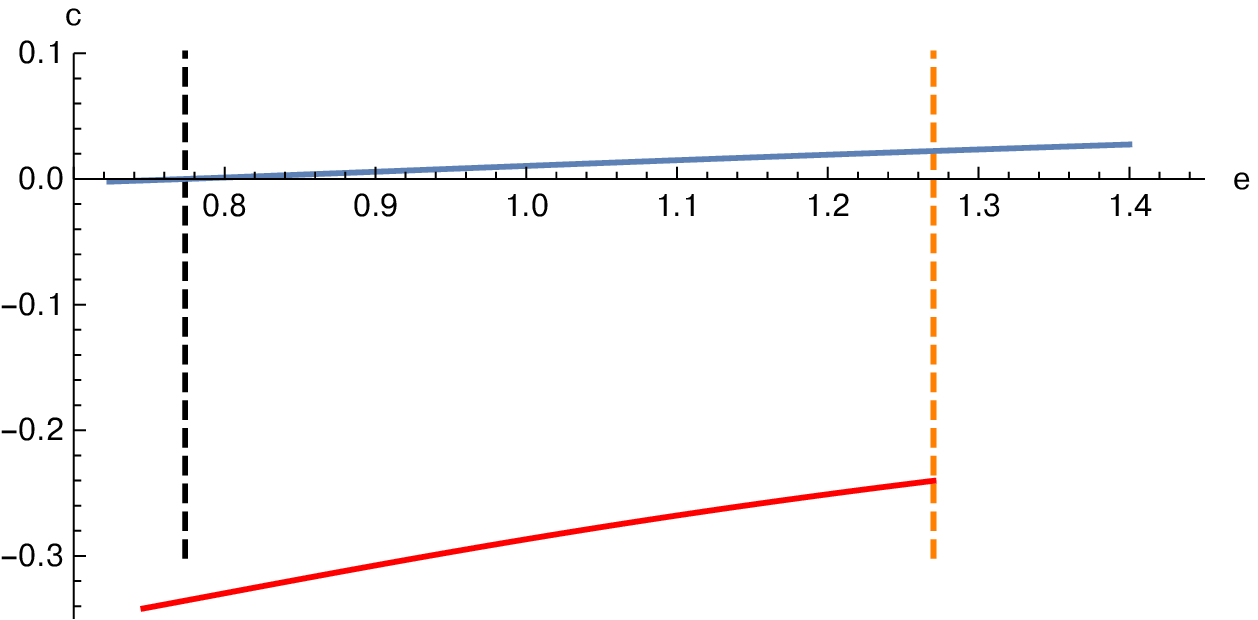}
\end{center}
  \caption{Phase diagram of the cascading gauge theory plasma in microcanonical ensemble. Solid blue lines represent
  deconfined, chirally symmetric phase of the theory --- the gravitational dual to these thermal states is a
  Klebanov-Tseytlin black hole. Solid red lines represent deconfined phase of the theory with the
  spontaneously broken chiral symmetry --- the gravitational dual to these thermal states is a Klebanov-Strassler black hole.
Left panel shows the entropy density versus the energy density
of the symmetric and the symmetry broken phases. The vertical dashed  orange line indicates the onset of the perturbative instability
of the chirally symmetric phase; the vertical dashed black line indicates the
onset of the instability of the sound waves in chirally symmetric phase of the
cascading plasma.
Right panel shows the speed of sound waves squared in symmetric and symmetry broken phases of the cascading plasma. 
 } \label{figure2}
\end{figure}

Microcanonical ensemble is relevant for dynamical questions (thermalization and equilibration) of gauge theory plasma.
Fig.~\ref{figure2} presents the phase diagram of the cascading gauge theory in microcanonical ensemble.
The solid blue curve denotes phase (B), and the solid red curve denotes phase (C).
Similar to \eqref{deffh} we introduced 
\begin{equation}
\hat{\cale}=16\pi G_5\ \cale\,,\qquad \hat{s}=16\pi G_5\ s\,.
\eqlabel{deffh2}
\end{equation}
A vertical orange line 
\begin{equation}
\hat{\cale}_{\csb}=1.270093(1)\ \Lambda^4
\eqlabel{or2}
\end{equation}
indicates the onset of the chiral symmetry breaking instability \cite{Buchel:2010wp}. 
Notice that here the phase (C) exists for $\cale\le {\cale}_{\csb}$ and dominates over
the phase (B). In other words, insisting on homogeneous and isotropic evolution,
the KSBH is the end point of the perturbative instability of the KTBH at sufficiently
low energy densities. Such a phenomenon in a context of {\it exotic} black holes
was identified in \cite{Buchel:2017map}.

The KSBH is both thermodynamically and dynamically unstable. The right panel of fig.~\ref{figure2}
presents the speed of sound waves as a function of the energy density for (B) (blue curve) and (C) (red curve)
phases of the cascading gauge theory plasma.
A vertical black line 
\begin{equation}
\hat{\cale}_{u}=0.723488\ \Lambda^4
\eqlabel{bl2}
\end{equation}
indicates the onset of the  perturbative instability in phase (B) associates with the breaking of the
translational invariance due to the condensation of the hydrodynamic sound waves \cite{Buchel:2009bh}.

\section{Landscape: KKLT de Sitter vacua in string theory}\label{lan}

\begin{figure}[t]
\begin{center}
\psfrag{e}{{$\frac{\hat{\cale}}{\Lambda^4}$}}
\psfrag{q3}{{$Q_b^{\rm D3}$}}
\psfrag{c}{{$c_s^2$}}
\includegraphics[width=5in]{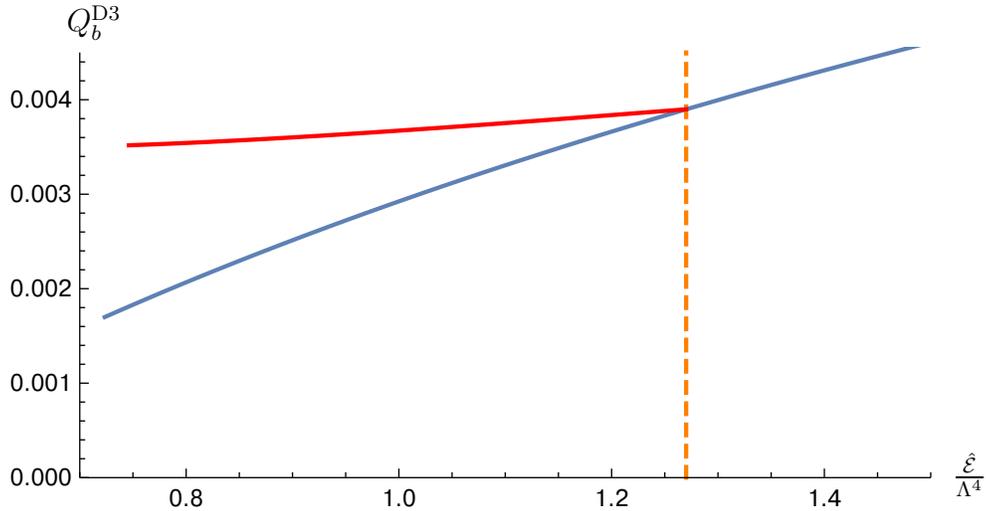}
\end{center}
  \caption{The Maxwell D3-brane charge of Klebanov-Tseytlin black holes  (solid blue line) and
Klebanov-Strassler black holes (solid red line).  Dashed  orange line indicates onset of the perturbative instability
of KT black holes, spontaneously breaking $U(1)\supset \zet_2$ chiral symmetry.
The end point of the evolution of the instability of a KT black hole for $\cale<\cale_{\csb}$,
under homogeneity and isotropy constraints of the horizon, is a KS black hole
at the corresponding energy density.
 } \label{figure3}
\end{figure}

Following \cite{Bena:2012ek}, we compute the Maxwell D3-brane charge of the conifold black hole horizons $Q_b^{\rm D3}$.
A negative value of the horizon charge would indicate that the anti-D3 brane singularity is physical, according to
classification \cite{Gubser:2000nd}. The results of the computations are presented in fig.~\ref{figure3}.
The solid blue curve represents the Maxwell charge of KTBH, and the solid red curve represents the charge of the
KSBH. Both charges are never negative; for $\cale<\cale_{\csb}$, when the KSBH has a higher entropy
then the corresponding energy density KTBH,
\begin{equation}
Q_b^{\rm D3}\bigg|_{KSBH}\ >\ Q_b^{\rm D3}\bigg|_{KTBH}\,.
\eqlabel{ktkscharge}
\end{equation}

\section{Conclusion}

In this paper we reported on the physical properties of Klebanov-Strassler black holes on warped deformed conifold with
fluxes. These black holes are important for understanding of the holographic correspondence for the confining
KS gauge theory \cite{Klebanov:2000hb}. They were also long-sought objects in the context of
KKLT de Sitter vacua constructions in string theory \cite{Kachru:2003aw}.

We established the existence of KSBHs. We determined that these black holes have a negative specific heat,
and are dynamically unstable due to the hydrodynamic sound wave fluctuations (breaking the homogeneity of the horizon).

KSBHs realize the homogeneous and isotropic deconfined phase of the cascading gauge theory plasma with spontaneously broken
chiral symmetry. While the corresponding phase never dominates in canonical ensemble, it has a higher entropy
density compare to the homogeneous and isotropic phase with unbroken chiral symmetry at the corresponding
energy density (the gravitational dual to KTBHs) below some critical energy density, \ie $\cale <\cale_{\csb}$.
Of course, since ultimately the homogeneity assumption is not valid (KSBHs are dynamically unstable),
the equilibrium states of the cascading gauge theory below $\cale_{\csb}$ remain unknown --- for sure they
can not be homogeneous and isotropic.

Finally, we demonstrated that KSBHs can not shield (conjectured) anti-D3 brane singularity
on the warped deformed conifold with fluxes.

\section*{Acknowledgments}
Research at Perimeter
Institute is supported by the Government of Canada through Industry
Canada and by the Province of Ontario through the Ministry of
Research \& Innovation. This work was further supported by
NSERC through the Discovery Grants program.

\appendix
\section{Technical details on construction of Klebanov-Strassler black hole}

Following \cite{Buchel:2010wp}, we present here technical details necessary to reproduce the results
reported in this paper. 

\subsection{Effective action and boundary asymptotics}
Five-dimensional effective action describing black holes on warped deformed
conifold with fluxes and $SU(2)\times SU(2)\times \zet_2$ global symmetry takes form:
\begin{equation}
\begin{split}
S_5=& \frac{108}{16\pi G_5} \int_{\calm_5} {\rm vol}_{\calm_5}\ \Omega_1 \Omega_2^2\om_3^2\ 
\biggl\lbrace 
 R_{10}-\frac 12 \left(\nabla \Phi\right)^2\\
&-\frac 12 e^{-\Phi}\left(\frac{(h_1-h_3)^2}{2\om_1^2\om_2^2\om_3^2}+\frac{1}{\om_3^4}\left(\nabla h_1\right)^2
+\frac{1}{\om_2^4}\left(\nabla h_3\right)^2\right)
\\
&-\frac 12 e^{\Phi}\left(\frac{2}{\om_2^2\om_3^2}\left(\nabla h_2\right)^2
+\frac{1}{\om_1^2\om_2^4}\left(h_2-\frac P9\right)^2
+\frac{1}{\om_1^2\om_3^4} h_2^2\right)
\\
&-\frac {1}{2\Omega_1^2\Omega_2^4\om_3^4}\left(4{\om}_0+ h_2\left(h_3-h_1\right)+\frac 19 P h_1\right)^2
\biggr\rbrace\,,\\
\end{split}
\eqlabel{5action}
\end{equation}
where $R_{10}$ is given by
\begin{equation}
\begin{split}
R_{10}=R_5&+\left(\frac{1}{2\om_1^2}+\frac{2}{\om_2^2}+\frac{2}{\om_3^2}-\frac{\om_2^2}{4\om_1^2\om_3^2}
-\frac{\om_3^2}{4\om_1^2\om_2^2}-\frac{\om_1^2}{\om_2^2\om_3^2}\right)-2\Box \ln\left(\om_1\om_2^2\om_3^2\right)\\
&-\biggl\{\left(\nabla\ln\om_1\right)^2+2\left(\nabla\ln\om_2\right)^2
+2\left(\nabla\ln\om_3\right)^2+\left(\nabla\ln\left(\om_1\om_2^2\om_3^2\right)\right)^2\biggr\}\,,
\end{split}
\eqlabel{ric5}
\end{equation}
where $R_5$ is the five dimensional Ricci scalar of the metric 
\begin{equation}
ds_{5}^2 =g_{\mu\nu}(y) dy^{\mu}dy^{\nu}\,,\qquad y^\mu=\{t,x_1,x_2,x_3\}\,.
\eqlabel{5met}
\end{equation}
The full ten-dimensional metric ansatz is
\begin{equation}
ds_{10}^2 =g_{\mu\nu}(y) dy^{\mu}dy^{\nu}+\om_1^2(y) g_5^2
+\om_2^2(y) \left[g_3^2+g_4^2\right]+\om_3^2(y) \left[g_1^2+g_2^2\right],
\eqlabel{10met}
\end{equation}
where
\begin{equation}
\begin{split}
&g_1=\frac{\a^1-\a^3}{\sqrt 2}\,,\qquad g_2=\frac{\a^2-\a^4}{\sqrt 2}\,,\\
&g_3=\frac{\a^1+\a^3}{\sqrt 2}\,,\qquad g_4=\frac{\a^2+\a^4}{\sqrt 2}\,,\\
&g_5=\a^5\,,
\end{split}
\eqlabel{3form1}
\end{equation}
\begin{equation}
\begin{split}
&\a^1=-\sin\theta_1 d\phi_1\,,\qquad \a^2=d\theta_1\,,\\
&\a^3=\cos\psi\sin\theta_2 d\phi_2-\sin\psi d\theta_2\,,\\
&\a^4=\sin\psi\sin\theta_2 d\phi_2+\cos\psi d\theta_2\,,\\
&\a^5=d\psi+\cos\theta_1 d\phi_1+\cos\theta_2 d\phi_2\,.
\end{split}
\eqlabel{3form2}
\end{equation}

We find it convenient to introduce
\begin{equation}
\begin{split}
h_1=&\frac 1P\left(\frac{K_1}{12}-36\Om_0\right)\,,\qquad h_2=\frac{P}{18}\ K_2\,,\qquad 
h_3=\frac 1P\left(\frac{K_3}{12}-36\Om_0\right)\,,\\
\Om_1=&\frac 13 f_c^{1/2} h^{1/4}\,,\qquad \Om_2=\frac {1}{\sqrt{6}} f_a^{1/2} h^{1/4}\,,\qquad 
\Om_3=\frac {1}{\sqrt{6}} f_b^{1/2} h^{1/4}\,.
\end{split}
\eqlabel{redef}
\end{equation}
The five-dimensional metric ansatz for the KSBH is
\begin{equation}
ds_5^2=H^{-1/2}\biggl(-(1-x)^2 dt^2+dx_1^2+dx_2^2+dx_3^2\biggr)+g_{xx}\ dx^2
\eqlabel{5dmet2}
\end{equation}
where
\begin{equation}
H(x)=(2x-x^2) h(x)
\eqlabel{defH}
\end{equation}
$x$ is the compactified radial coordinate: 
\begin{equation}
x\in (0,1)
\eqlabel{rangex}
\end{equation}
From the effective action \eqref{5action} we obtained non-linear system of ODEs for
8 functions  (see \cite{Buchel:2010wp} for the explicit form of the system): 
\begin{itemize}
\item the three flux functions: $\{K_1,K_2,K_3\}$;
\item the overall warp factor of deformed $T^{1,1}$: $\{h\}$;
\item the three deformation warp factors "inside'' $T^{1,1}$: $\{f_a,f_b,f_c\}$;
\item the string coupling constant: $\{g\}$
\end{itemize}
All equations of motion are second order in the radial derivative $\frac{d}{dx}$ (the second derivatives
enter linearly the system), thus we need $8\times 2=16$ integration constants to specify a
numerical solution.  

The UV asymptotic corresponds to $x\to 0_+$. We find:
\begin{equation}
K_1=P^2 g_0 k_s-\frac12 P^2 g_0\ \ln x+\sum_{n=3}^\infty\sum_k\ k_{1nk}\ x^{n/4}\ \ln^k x\,,
\eqlabel{uv1}
\end{equation}
\begin{equation}
K_2=1+\sum_{n=3}^\infty\sum_k\ k_{2nk}\ x^{n/4}\ \ln^k x\,,
\eqlabel{uv2}
\end{equation}
\begin{equation}
K_3=P^2 g_0 k_s-\frac12 P^2 g_0\ \ln x+\sum_{n=3}^\infty\sum_k\ k_{3nk}\ x^{n/4}\ \ln^k x\,,
\eqlabel{uv3}
\end{equation}
\begin{equation}
f_a=a_0\biggl(1+\sum_{n=3}^\infty\sum_k\ f_{ank}\ x^{n/4}\ \ln^k x\biggr)\,,
\eqlabel{uv4}
\end{equation}
\begin{equation}
f_b=a_0\biggl(1+\sum_{n=3}^\infty\sum_k\ f_{bnk}\ x^{n/4}\ \ln^k x\biggr)\,,
\eqlabel{uv5}
\end{equation}
\begin{equation}
f_c=a_0\biggl(1+\sum_{n=2}^\infty\sum_k\ f_{cnk}\ x^{n/4}\ \ln^k x\biggr)\,,
\eqlabel{uv6}
\end{equation}
\begin{equation}
h=\frac{P^2g_0}{a_0^2}\left(\frac 18+\frac {k_s}{4}\right)-\frac  {P^2 g_0}{8a_0^2}\ \ln x+\sum_{n=2}^\infty\sum_k\ h_{nk}\ x^{n/4}\ \ln^k x\,,
\eqlabel{uv7}
\end{equation}
\begin{equation}
g=g_0\left[\ 1+\sum_{n=2}^\infty\sum_k\ g_{nk}\ x^{n/2}\ \ln^k x \ \right]\,.
\eqlabel{uv8}
\end{equation}
The expansion depends on 4 microscopic parameters
\begin{equation}
\{P^2\,, g_0\,, a_0\,, k_s\}\,,
\eqlabel{par}
\end{equation}
\begin{itemize}
\item $P^2 g_0$ --- the dimensionless parameter of the cascading theory
(which must be large for the gravity approximation to be valid);
\item $a_0^2=4\pi G_5\ sT$ --- where $s$ is the entropy density
and $T$ is the temperature of KSBH;
\item the strong coupling scale $\Lambda$ of the cascading
gauge theory is given by
\begin{equation}
k_s\equiv \frac 12\ \ln \left(\frac{a_0^2}{\Lambda^4}\right)=\frac 12\ \ln \left(\frac{4\pi G_5 sT}{\Lambda^4}\right)\,.
\eqlabel{defl}
\end{equation}
\end{itemize}
Besides \eqref{par}, the expansions \eqref{uv1}-\eqref{uv8} are characterized by 7 expectation values:
\nxt those of dimension-3 operators:
\begin{equation}
\left\{{ df_{0}}\equiv \frac12 \left(f_{a30}-f_{b30}\right)\,,\ {
dk_{10}}\equiv \frac 12\left(k_{130}-k_{330}\right) \right\}\,,
\eqlabel{vev3}
\end{equation}
\nxt those of dimension-4 operators:
\begin{equation}
\{ f_{a40}\,,\  g_{40} \}\,,
\eqlabel{vev4}
\end{equation}
\nxt that of a dimension-6 operator:
\begin{equation}
\{f_{a60} \}\,,
\eqlabel{vev6}
\end{equation}
\nxt that of a dimension-7 operator:
\begin{equation}
\{k_{270}\}\,,
\eqlabel{vev7}
\end{equation}
\nxt and finally, that of a dimension-8 operators:
\begin{equation}
\{f_{a80} \}\,.
\eqlabel{vev8}
\end{equation}

Introducing $y=1-x$, the regular horizon $y\to 0_+$ asymptotics of 
$$\{K_1,\ K_2,\ K_3,\ f_a,\ f_b,\ f_c,\ h,\ g\}\,,$$ take form:
\begin{equation}
\begin{split}
&K_i=\sum_{n=0}^\infty k_{ihn}\ y^{2n}\,,\qquad i=1,2,3\,,\\
&f_{\alpha}=a_0\ \sum_{n=0}^\infty f_{\a hn}\ y^{2n}\,,\qquad \a=a,b,c\,,\\
&h=\sum_{n=0}^\infty h_{hn}\ y^{2n}\,,\qquad g=g_0\ \sum_{n=0}^\infty g_{hn}\ y^{2n}\,.
\end{split}
\eqlabel{ir1}
\end{equation} 
Here, the expansion is characterized by 9 parameters:
\begin{equation}
\begin{split}
&\biggl\{k_{1h0}\,, k_{2h0}\,, k_{3h0}\,,f_{ah0}\,,  f_{bh0}\,,  f_{ch0}\,, f_{ch1}\,,
h_{h0}\,, g_{h0}\biggr\}\,.
\end{split}
\eqlabel{ircond}
\end{equation}

Equations of motion are invariant under the following symmetries:
\nxt SYM-A: 
\begin{equation} 
h\to \lambda^{-2}\ h\,,\qquad  f_{a,b,c}\to \lambda\ f_{a,b,c}\,,
\eqlabel{scaling1}
\end{equation}
\nxt SYM-B: 
\begin{equation} 
P\to \lambda^{-1}\ P\,,\qquad  g\to \lambda^2\ g\,,
\eqlabel{scaling2}
\end{equation}
\nxt SYM-C: 
\begin{equation} \eqlabel{scaling3}
h\to \lambda^{2} h\,,\ f_{a,b,c} \to f_{a,b,c}\,,\ K_{1,3}\to
\lambda^2 K_{1,3}\,,\ K_2\to K_2\,,\ g\to g\,,\ P\to \lambda P
\end{equation}
Above rescaling symmetries can be used to set:
\begin{equation}
P=g_0=a_0=1
\eqlabel{setmicro}
\end{equation}
It is important to present physical results in dimensionless form --- from \eqref{defl} we see that
with \eqref{setmicro}
\begin{equation}
\Lambda=e^{-k_s/2}
\eqlabel{setL}
\end{equation}

\subsection{Holographic renormalization, thermodynamic quantities and horizon $D3$ Maxwell charge}

Holographic renormalization of the theory \eqref{5action} with unbroken $U(1)$ symmetry was implemented
in \cite{Aharony:2005zr}. To study KSBH we need renormalization with $U(1)\supset \zet_2$. For quantities
of interest, this is easily done
with the following substitutions:
\begin{equation}
K^{KT}=\frac 12 K_1+\frac 12 K_3\,,\qquad \Om_1^{KT}=3\Om_1\,,\qquad \Om_2^{KT}=\frac{\sqrt{6}}{2}
\left(\Om_2+\Om_3\right)
\eqlabel{subs}
\end{equation}

For relevant thermodynamic quantities we find:
\begin{itemize}
\item the temperature $T$:
\begin{equation}
\begin{split}
&\frac{T}{\Lambda}=\frac{e^{k_s/2}}{16 \pi} \biggl(-\frac{1}{f_{ah0}^2 f_{bh0}^2\ g_{h0}\ h_{h0}^2\ (f_{ch0}+2 f_{ch1})}
\biggl(
(16\ f_{ah0}^2\ k_{2h0}^2\\
&+16\ f_{bh0}^2\ k_{2h0}^2-64\ f_{bh0}^2\ k_{2h0}+64\ f_{bh0}^2)\ g_{h0}^2
-16\ f_{ah0}\ f_{bh0}\ h_{h0}\ (\\
&3\ f_{bh0}-3\ f_{ah0}+4\ f_{ch0})\ (4\ f_{ch0}-3\ f_{bh0}+3\ f_{ah0})\ g_{h0}
\\
&+18\ f_{ah0}\ f_{bh0}\ k_{1h0}^2-36\ f_{ah0}\ f_{bh0}\ k_{1h0}\ k_{3h0}+18\ f_{ah0}\ f_{bh0}\ k_{3h0}^2\biggr)
\biggr)^{1/2}
\end{split}
\eqlabel{TL}
\end{equation}
\item the entropy density $s$:
\begin{equation}
16\pi G_5\ \frac{s}{\Lambda^3}=4\pi\ e^{3 k_s/2}\  h_{h0}^{1/2}\ f_{ch0}^{1/2}\ f_{ah0}\ f_{bh0}
\eqlabel{sL3}
\end{equation}
\item the energy density $\cale$:
\begin{equation}
16\pi G_5\ \frac{\cale}{\Lambda^4}=\ e^{2 k_s}\ (3 -12\ f_{a40})
\eqlabel{eL4}
\end{equation}
\item the free energy density $\calf$:
\begin{equation}
{\calf}=\cale- s\ T
\eqlabel{fL4}
\end{equation}
\end{itemize}
A highly nontrivial check is a numerical verification of the first law of the thermodynamics. We find (for constant
$\Lambda$)
\begin{equation}
\bigg|1- \frac{T ds}{d\cale}\bigg|_{{\rm numerically}}=\bigg|\ 1- \frac{\frac{T}{\Lambda} d\left(s/\Lambda^3\right)}
{d\left(\cale/\Lambda^4\right)}\ \bigg|_{{\rm numerically}}\ \sim\ 5\times 10^{-6}\ \cdots\ 3\times 10^{-5}
\eqlabel{check}
\end{equation}

From \cite{Aharony:2007vg}
\begin{equation}
16\pi G_5=\frac{16\pi G_{10}
}{{\rm vol} \left(T^{1,1}\right)}=\frac{27}{16\pi^3}\ 16\pi G_{10}=
216\pi^4(\a')^4g_0^2
\eqlabel{normf}
\end{equation}

From \cite{Bena:2012ek}
the Maxwell charge at the black hole horizon is
\begin{equation}
Q^{{\rm D3}}_b=\frac{1}{54\pi}\ \biggl(k_{1h0}\ (2-k_{2h0})+k_{2h0}\ k_{3h0}\biggr)
\eqlabel{d3charge}
\end{equation}

\subsection{Numerical procedure}

We are now ready to formulate our numerical procedure, and count the parameters of the 
solution:
\nxt We integrate the differential equations along $x$-coordinate
\begin{equation}
0\le x \le 1\,,
\eqlabel{xrange}
\end{equation}
with $x=0$ being the boundary and $x=1$ being the horizon. 
\nxt We use various scaling symmetries discussed above to set \eqref{setmicro}.
\nxt Altogether we need to integrate 8 functions 
\begin{equation}
\{K_1,K_2,K_3,f_a,f_b,f_c,h,g\}\,,
\eqlabel{functions}
\end{equation}
for a given set of the remaining microscopic parameter 
$\{k_s\}$. 
\nxt The solution is  then determined by 7 UV parameters \eqref{vev3}-\eqref{vev8}, and 9 IR parameters \eqref{ircond}:  
\begin{equation}
\begin{split}
{\rm UV}:&\qquad \{df_{0}\,, dk_{10}\,, f_{a40}\,, g_{40}\,, f_{a60}\,, k_{270}\,, f_{a80}\}\,,\\
{\rm IR}:&\qquad \{k_{1h0}\,, k_{2h0}\,, k_{3h0}\,, f_{ah0}\,, f_{bh0}\,, f_{ch0}\,, f_{ch1}\,,  h_{h0}\,, g_{h0}\}\,.
\end{split}
\eqlabel{paruvir}
\end{equation} 
Overall we have 16 parameters, precisely what is necessary to determine \eqref{functions} from the appropriate second 
order differential equations.

We follow numerical method introduced in \cite{Aharony:2007vg}. In a nutshell, for a fixed  
microscopic parameter $\{k_s\}$, we choose a 'trial' set of parameters \eqref{paruvir}
and integrate (a double set of) the equations of motion for \eqref{functions} from the 
UV ($x_{initial}=0.001$)  to $x=0.5$, and from the IR ($y_{initial} =0.001$) to $y=0.5$. 
A solution \eqref{paruvir} of the boundary value problem implies that the mismatch vector
\begin{equation}
\begin{split}
\vec{v}_{mismatch}\equiv& \biggl(K_1^b-K_1^h\,,\ (K_1^b+K_1^h)'\,,\ K_2^b-K_2^h\,,\ (K_2^b+K_2^h)'\,,\ 
K_3^b-K_3^h\,,\ \\
&(K_3^b+K_3^h)'\,,\ f_a^b-f_a^h\,,\ (f_a^b+f_a^h)'\,,\ f_b^b-f_b^h\,,\ (f_b^b+f_b^h)'\,,\ 
f_c^b-f_c^h\,,\ \\
&(f_c^b+f_c^h)'\,,\ h^b-h^h\,,\ (h^b+h^h)'\,,\ g^b-g^h\,,\ (g^b+g^h)'\biggr)_{x=y=0.5}\,,
\end{split}
\eqlabel{mismatch}
\end{equation} 
with the superscripts $\ ^b$ and $\ ^h$ referring to the boundary (UV) and the horizon (IR) 
integrations, vanishes. At each iteration we adjust the  set of parameters \eqref{paruvir}
along the direction of the steepest decent for 
$||\vec{v}_{mismatch}||$. In practice, for a valid numerical solution we were able to  
achieve 
\begin{equation}
||\vec{v}_{mismatch}||\ \sim\  10^{-37}\cdots 10^{-32}\,.
\eqlabel{normachive}
\end{equation}

\end{document}